# Non-perturbative generation of above-threshold harmonics from pre-excited argon atoms in intense mid-infrared laser fields

Guihua Li[1], Hongqiang Xie[1,2], Ziting Li[1], Jinping Yao[1], Bin Zeng[1], Wei Chu[1], and Ya Cheng[1,*]

[1]*State Key Laboratory of High Field Laser Physics, Shanghai Institute of Optics and Fine Mechanics, Chinese Academy of Sciences, Shanghai 201800, China*
[2]*University of Chinese Academy of Sciences, Beijing 100049, China*
*Corresponding author: ya.cheng@siom.ac.cn*

**We experimentally investigate the generation of above-threshold harmonics completely from argon atoms on an excited state using mid-infrared femtosecond laser pulses. The highly nonlinear dependences of the observed signal on the pulse energy and polarization of the driver laser pulses indicate its non-perturbative characteristic.**

High-order harmonic generation (HHG) from atoms and molecules driven by intense laser pulses has attracted significant attention in the last three decades [1-3]. As a non-perturbative nonlinear process, HHG not only opens a new route to generate coherent tabletop x-ray sources and attosecond pulses [4-6], but also provides an effective tool for probing the ultrafast dynamics of atoms/molecules on angstrom-scale [7-9]. High-order harmonic generation can be well understood in the framework of semiclassical three-step model [10,11]: (i) tunnel ionization of the electrons into the continuum state; (ii) generation of recolliding electrons moving towards the parent ion in the oscillating laser field; and (iii) recombination of the returned electrons with the ions that leads to emission of coherent photons with photon energies above the Coulomb potential.

Recently, high-order harmonic generation from excited states of atoms/molecules has been attracting growing interest due to its potential to generate high-order harmonics with enhanced conversion efficiency and/or extended cut-off energy [12-14]. Besides, it also provides us an opportunity to investigate extreme nonlinear optical process (i.e., non-perturbative characteristics) of excited states, which is fundamental in photochemistry and photobiology. However, most investigations on HHG from excited states are based on a coherent superposition of ground state and excited states because of the strong laser fields involved, which enable generation of harmonics from both the ground and excited states [15-17]. Such scheme is beneficial for promoting the conversion efficiency of HHG. On the other hand, around one decade ago, Paul et al. experimentally demonstrated HHG completely from the excited states of alkali-metal atoms (i.e., rubidium), which was realized by cascade excitation in the presence of a weak cw laser field [18]. It should be noted that HHG from pure excited states of noble gases has not been reported, as a similar near-resonant optical excitation requires strong sources of extreme ultraviolet light [14,19,20]. Another difficulty is that for atoms and molecules on the excited states, their ionization potentials are too low to maintain a sufficiently low Keldysh parameter (i.e., $\gamma = \sqrt{I_p/2U_p}$, where $I_p$ represents ionization potential and $U_p$ represents ponderomotive energy) at the wavelengths near 800 nm [21]. This issue can be nicely overcome because of the rapid development of intense mid-infrared laser sources whose extended wavelengths can ensure the ionization of the excited atoms/molecules in the tunnel regime [22].

In this work, we experimentally generate the above-threshold harmonics completely from pre-excited argon atoms using a simple all-optical approach. By comparing the characteristics of an above-threshold harmonic and that of a below-threshold harmonic, our results reveal the non-perturbative physical picture of the above-threshold harmonics.

The pump-probe experimental setup is schematically illustrated in Fig. 1, which is similar to that used in our previous works [23-25]. Briefly, the mid-infrared femtosecond laser pulses were provided by optical parametric amplifier (OPA, HE-TOPAS, Light conversion, Ltd.) pumped by a commercial Ti:sapphire laser system (Legend Elite-Duo, Coherent Inc.). In this experiment, a signal beam at the wavelength of 1420 nm was used as the pump and the idler beam at the wavelength of 1820 nm was served as the probe. The maximum energy of pump pulse was ~1 mJ. The probe pulse energy can be continuously adjusted from 640 $\mu$J to 2 $\mu$J using a variable neutral density filter (VAF). The diameter of probe beam was reduced by a factor of 2.5 with the help of a telescope system inserted in the probe beam path. In this case, the pump energy was high enough to generate a 13-mm-long filament, whereas the probe laser alone cannot form any visible plasma owing to its low pulse energy. The pump pulse was set to be circularly polarized for minimizing the supercontinuum generation in the vicinity of the 3rd and 7th harmonics. Meanwhile, the polarization of the probe pulse can be continuously changed from linear to circular by rotating a quarter-wave plate (QWP2). The time delay between the pump and probe pulses was controlled using a delay line in the pump beam path. After being combined by a dichroic mirror (DM1), the pump and the probe pulses were collinearly focused into a gas chamber filled with argon by a 15-cm focal-length lens. The gas pressure of argon was fixed at 1000 mbar. The generated harmonics after the chamber were recorded using a 1200-grooves/mm grating spectrometer (Andor Shamrock 303i) after being focused with a 10-cm focal-length lens onto the entrance slit.

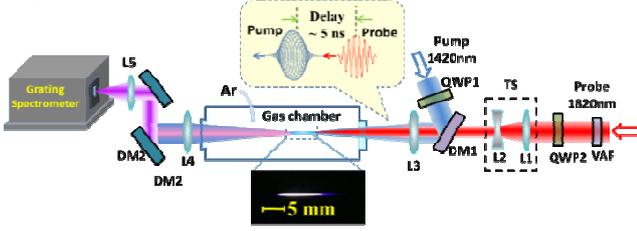

**Fig. 1.** Schematic diagram of the experimental setup. VAF: variable attenuation filter; QWP: quarter-wave plate; DM: dichroic mirror; TS: telescope system.

As shown in Fig. 2(a) and (b), strong 3rd and 7th harmonics of the probe pulse can be generated when both the pump and probe pulses were focused into the gas chamber filled with argon gas. Here, the probe pulse energy was fixed at ~70 μJ. The time delay between the pump and probe pulse was adjusted to be ~5 ns; and the probe pulse was chosen to be linearly polarized. In addition, it can be seen that no significant harmonic signal can be observed when either of the two laser beams was blocked.

The reason to choose a time delay of ~5 ns for achieving the strongest harmonic signal is indicated in Fig. 2(c), in which we present the measured 7th harmonic signal intensity as a function of the time delay between the pump and probe pulses. It is clear that 7th harmonic signal starts to appear at ~2 ns after the pump pulse and reaches its peak intensity around ~5 ns and then shows a gradual decay with the increasing time delay. As we will show later, the temporal dependence we observe in Fig. 2(c) directly reflects the dynamics of excitation of the argon atoms with the intense pump pulses.

Figure 2(d) shows the measured spatial beam profiles of 3rd (red squares) and 7th (black circles) harmonics. It can be seen that both the beam profiles of 3rd and 7th harmonics are nearly Gaussian. We fit the experimental data with Gaussian functions and estimate the divergence angles of 3rd and 7th harmonics, which were 13 mrad and 8 mrad in our experiments, respectively. Besides, the conversion efficiency of 7th harmonics from the excited argon atoms is measured to be ~$10^{-7}$.

Figure 3(a) shows the measured 3rd (red squares) and 7th harmonic (black circles) signals from the excited-state argon atoms as a function of the driver pulse energy. In this measurement, the diameter of the incident probe beam was first reduced by a factor of 2.5 with a telescope system. The pump pulse energy was fixed at 1 mJ and the probe pulse energy was continuously adjusted using the neutral density filter. It can be seen that the measured 3rd harmonic signal dramatically increases with the increasing probe laser energy, which can be well fitted with $y = ax^3$, i.e., it perfectly follows the $I^3$ law [26,27]. On the other hand, the 7th harmonic signal starts to appear when the probe pulse energy reaches 48 μJ. The 7th harmonic signal then grows nearly linearly with the increasing driver pulse energy.

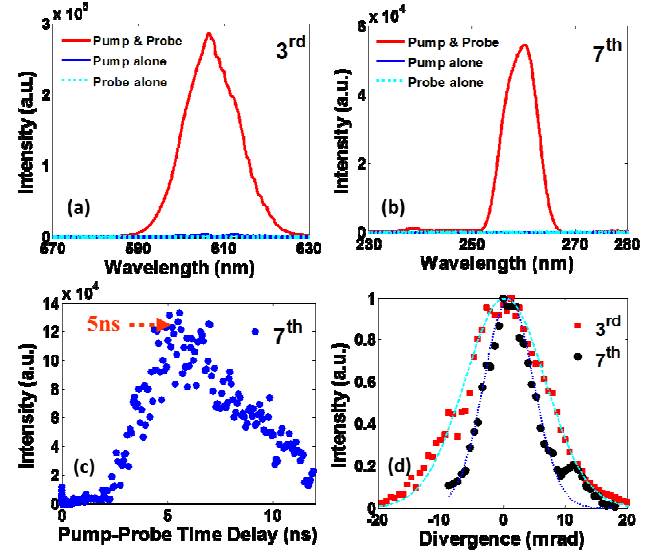

**Fig. 2.** Typical spectra of 3rd harmonic (a) and 7th harmonic (b); (c) measured 7th harmonic signal as a function of pump-probe time delay; (d) measured beam profiles of 3rd harmonic (red squares) and 7th harmonic (black circles).

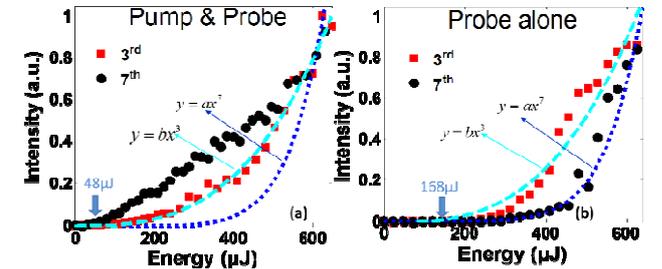

**Fig. 3.** (a) Measured 7th (black circles) and 3rd (red squares) harmonic signals from excited argon atoms as a function of the probe pulse energy; (b) measured 7th and 3rd harmonic signals from unexcited argon atoms as a function of probe laser energy. The green dashed and blue dotted curves are the corresponding fitting curves with function $I^q$, where $q$ is the harmonic order.

For comparison, we also generated the harmonics from the ground state of argon and measured their signal intensities as a function of the driving laser energy, as shown in Fig. 3(b). In this measurement, we blocked the pump pulse, thus all the argon atoms remained in the ground state. It is worth mentioning that in this case, to generate the 3rd and 7th harmonics by the probe pulse alone, we removed the telescope system in the probe beam path to enhance the peak intensity in the focus region by almost a factor of 6. Both the 3rd and 7th harmonic signals increase with the increasing driver pulse energy. Likewise, we fit the measured data of 3rd (red squares) and 7th (black circles) harmonics from the ground-state argon atoms with the function $I^q$, where $q$ is the corresponding harmonic order, and then found that both the 3rd and 7th harmonic signals show a fairly well agreement with the $I^q$ law [26,27].

The dependences of the 3rd harmonic and 7th harmonic signals from the pre-excited argon atoms on the ellipticity of polarization of the driver laser pulse are presented in Fig. 4. The ellipticity is defined as the ratio of the minor axis to the major axis of the ellipse, i.e., $\varepsilon = |E_x/E_y|$, where $E_x$ and $E_y$ represent the laser electric field strengths along the horizontal and vertical directions, respectively. To eliminate the influence of detection system on this measurement, we collected the 3rd and 7th harmonic signals using an integrating sphere and a fiber. As shown in Fig. 4, both the 3rd (red squares) and 7th (black circles) harmonic signals reach their maxima when the probe pulse is linearly polarized and gradually decreases when the probe pulse varies from linear polarization to circular polarization. The 7th harmonic signal becomes too weak to be detectable when the ellipticity increased to 0.7, whereas the 3rd harmonic signal only decreases by nearly one order of magnitude at the same ellipticity.

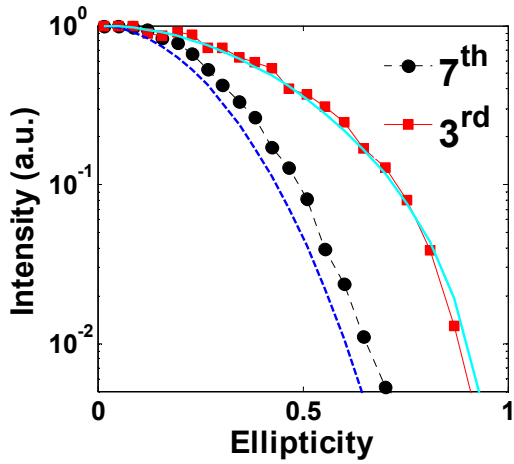

**Fig. 4.** (a) Measured dependences of 7th (black circles) and 3rd (red squares) harmonic signals from excited argon atoms on the elliptical polarization of the probe pulse. The green solid and blue dotted curves represent the theoretical predictions.

In the pump-probe experiments, we would like to stress that the probe pulse energy (i.e., 70 μJ) is too weak to generate any detectable harmonic signals from the ground state of argon. Moreover, because the time delay between the pump and probe pulses is on the order of a few nanoseconds, the harmonic generation contributed by the plasma effect can be excluded due to the relatively short lifetime of the plasma in the femtosecond laser filament [28]. Previously, it has been demonstrated that a large population of excited argon atoms can be prepared following the femtosecond pulse excitation, and then the population inversion between the excited states $C^3\Pi_u$ and $B^3\Pi_g$ of N2 molecules can be realized [29,30]. Therefore, it is reasonable to attribute the strong 3rd and 7th harmonic signal generated in our experimental conditions to the nonlinear interaction of the probe beam with the excited-state argon atoms prepared by the intense pump laser pulses. Besides, the temporal evolution of 7th harmonic signal in Fig. 2(c) agrees well with our previous observations, providing quantitative evidence on the role of the excited state of argon, i.e., [$Ar^*(4\,^3P_2)$] [30]. It is noticed that the photon energy of the 3rd harmonic signal is ~2 eV, whereas the photon energy of 7th harmonic is ~4.8 eV. The latter is higher than the ionization energy (~4.15 eV) of argon on the excited state [$Ar^*(4\,^3P_2)$]. In general, the above-threshold harmonic is a signature of the non-perturbative nonlinear process, which is further supported by the experimental results in Fig. 3. One can clearly see that the 3rd harmonic signal generated from either the excited state [Fig. 3(a)] or the ground state [Fig. 3(b)] can be well explained by the perturbative theory. In contrast, the 7th harmonic signal generated from the excited argon with the pump-probe scheme shows large deviations from the $I^7$ law predicted by the perturbation theory. Instead, a nearly linear dependence of the 7th harmonic on the intensity of the probe beam is observed. Furthermore, we have noticed that the threshold intensity for generating the 7th harmonics from the ground-state argon atoms is ~21 times as high as that from the pre-excited argon atoms.

An additional evidence to support the non-perturbative picture for the above-threshold harmonic is the observed strong dependence of the above-threshold harmonic (i.e., 7th harmonic) on the ellipticity of the driving laser pulses. In the perturbative regime, the dependence of harmonics on the ellipticity of the polarization of the pump pulses can be expressed by $I_q = \left[(1-\varepsilon^2)/(1+\varepsilon^2)\right]^{q-1}$ [27]. Using the expression, we calculate the ellipticity dependence of both the 3rd and 7th harmonics, which are shown as the green solid and blue dashed curves in Fig. 4. The experimental result and the prediction from the perturbative theory agrees very well for the 3rd harmonic signal. However, for the above-threshold harmonic (i.e., 7th harmonic) generation, the deviation between the experimental result and perturbative theory implies again that the 7th harmonic generation should originate from the non-perturbative process rather than the perturbative process. In the three-step model of HHG, the ellipticity dependence is a result of the diffusive wavepacket dynamics of the electrons in the combined Coulomb and laser fields, which will be investigated in the future.

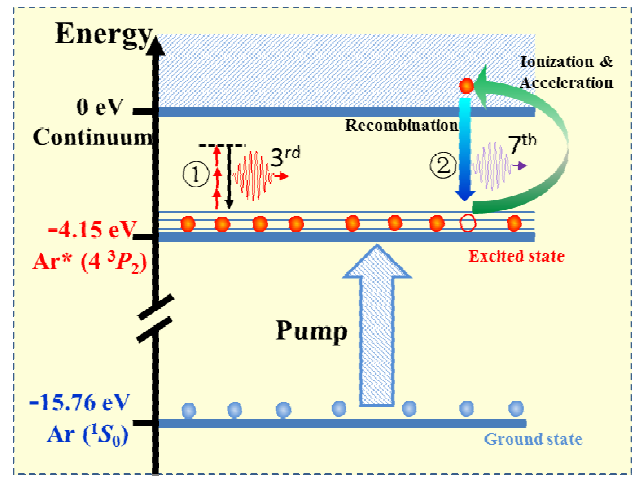

**Fig. 5.** The physical picture of the 3rd and the 7th harmonic generation from the excited argon atoms.

Based on the experimental results and discussions mentioned above, we provide the physical picture for the observed harmonics generated from the excited-state argon atoms, which is shown in

Fig. 5. In the first step, the pump pulses efficiently excite the argon atoms into a long-life metastable state $Ar^*(4\,^3P_2)$, which is caused by a two-step kinetic process including of three-body collisions and dissociation recombination. For the reasons mentioned above, the contribution from the ground state of argon to the HHG is negligible in the pump-probe measurements. The 3rd harmonic generation from the excited-state argon atoms is obviously a perturbative nonlinear process. In contrast, a non-perturbative process (labeled as ②) is suggested to be responsible for the above-threshold harmonic generation, i.e., an electron from the excited state is ionized by the probe laser pulses, and then finally recombined with parent ion after being accelerated and driven back in the oscillating laser field.

In conclusion, we have experimentally investigated the above-threshold (i.e., 7th) harmonic generated from the excited argon atoms using mid-infrared femtosecond laser pulses. By comparing the dependences of the above-threshold and below-threshold harmonic signals on the driving laser parameters, our results demonstrate that the below-threshold harmonic generation can be well understood in the framework of the perturbative theory, whereas the above-threshold harmonics should be generated essentially through the non-perturbative three-step process. Our work motivates the investigations on various extreme nonlinear optical processes in the excited atoms/molecules, opening the possibility for obtaining the spatial and temporal information of excited atoms/molecules with resolutions on the atomic scales.